\documentstyle[12pt,epsf]{article}
\newlength{\dinwidth}
\newlength{\dinmargin}
\newlength{\extraspace}
\newlength{\extraspaces}
\newcommand{\be}{\begin{equation}
\addtolength{\abovedisplayskip}{\extraspaces}
\addtolength{\belowdisplayskip}{\extraspaces}
\addtolength{\abovedisplayshortskip}{\extraspace}
\addtolength{\belowdisplayshortskip}{\extraspace}}
\newcommand{\ee}{\end{equation}}
\newcommand{\bdm}{\begin{displaymath}
\addtolength{\abovedisplayskip}{\extraspaces}
\addtolength{\belowdisplayskip}{\extraspaces}
\addtolength{\abovedisplayshortskip}{\extraspace}
\addtolength{\belowdisplayshortskip}{\extraspace}}
\newcommand{\edm}{\end{displaymath}}
\renewcommand{\thefootnote}{\fnsymbol{footnote}}
\def\simlt{\mathrel{\lower2.5pt\vbox{\lineskip=0pt\baselineskip=0pt
           \hbox{$<$}\hbox{$\sim$}}}}
\def\simgt{\mathrel{\lower2.5pt\vbox{\lineskip=0pt\baselineskip=0pt
           \hbox{$>$}\hbox{$\sim$}}}}
\catcode`@=11
\newcount\@tempcntc
\def\@citex[#1]#2{\if@filesw\immediate\write\@auxout{\string\citation{#2}}\fi
  \@tempcnta\z@\@tempcntb\m@ne\def\@citea{}\@cite{\@for\@citeb:=#2\do
    {\@ifundefined
       {b@\@citeb}{\@citeo\@tempcntb\m@ne\@citea\def\@citea{,}{\bf ?}\@warning
       {Citation `\@citeb' on page \thepage \space undefined}}%
    {\setbox\z@\hbox{\global\@tempcntc0\csname b@\@citeb\endcsname\relax}%
     \ifnum\@tempcntc=\z@ \@citeo\@tempcntb\m@ne
       \@citea\def\@citea{,}\hbox{\csname b@\@citeb\endcsname}%
     \else
      \advance\@tempcntb\@ne
      \ifnum\@tempcntb=\@tempcntc
      \else\advance\@tempcntb\m@ne\@citeo
      \@tempcnta\@tempcntc\@tempcntb\@tempcntc\fi\fi}}\@citeo}{#1}}
\def\@citeo{\ifnum\@tempcnta>\@tempcntb\else\@citea\def\@citea{,}%
  \ifnum\@tempcnta=\@tempcntb\the\@tempcnta\else
   {\advance\@tempcnta\@ne\ifnum\@tempcnta=\@tempcntb \else \def\@citea{--}\fi
    \advance\@tempcnta\m@ne\the\@tempcnta\@citea\the\@tempcntb}\fi\fi}
\catcode`@=12

\newcommand{\THDM}{Two-Higgs-Doublet Model}

\newcommand{\Ma}{M_{A^{0}}}
\newcommand{\Mg}{M_{H^{\pm}}}

\begin{document}
\begin{titlepage}
\begin{flushright}
BUHEP 95-6\\
hep-ph/9502343\\
February 17, 1995 \\
\end{flushright}
\vspace{24mm}
\begin{center}
\Large{{\bf The LHC Phenomenology of the CP-odd Scalar
in Two-Doublet Models}}
\end{center}
\vspace{5mm}
\begin{center}
Dimitris Kominis\footnote{Talk presented at the Conference {\it Beyond
the Standard Model IV}, Lake Tahoe, California, December 13-18, 1994.}
\footnote{e-mail address: kominis@budoe.bu.edu}
\\*[3.5mm]
{\normalsize\it Dept. of Physics, Boston University, 590 Commonwealth
Avenue,}\\
{\normalsize\it Boston, MA 02215}
\end{center}
\vspace{2cm}
\thispagestyle{empty}
\begin{abstract}
We discuss possible signatures of the $CP$-odd scalar of
two-doublet models at the LHC. We find that the inclusive two-photon
decay mode and the decay sequence $A^0 \rightarrow Zh, h \rightarrow
\gamma \gamma$, where $h$ is a $CP$-even neutral scalar, can give viable
signals in fairly large and complementary regions of parameter space.
\end{abstract}
\end{titlepage}
\newpage

\renewcommand{\thefootnote}{\arabic{footnote}}
\setcounter{footnote}{0}
\setcounter{page}{2}

The prospects of detecting the Higgs bosons
of non-minimal models at future high-energy colliders
have been the focus of many studies.
In the \THDM\ (THDM), the simplest 
extension of the scalar sector of the Standard
Model, the masses and self-couplings of the Higgs bosons are unknown
parameters.  Nonetheless,
restrictions on their range of values can be imposed from
existing experimental data, as well as from considerations
of triviality and stability of the effective
potential \cite{dn,me}. The purpose of the study reported here is to
determine the potential of the Large Hadron Collider to probe the $CP$-odd
scalar particle of the THDM in the region of parameter space
defined by the triviality (and stability) bounds.

We shall consider a THDM with an exact discrete symmetry
meant to ensure natural flavor conservation \cite{gw}. This
symmetry is commonly implemented in one of two ways, which differ in the
manner the fermions couple to the scalar doublets, and are referred to as
Model~I and Model~II in the literature \cite{hhg}. After
the two doublets
acquire vacuum expectation values $v_1, v_2$, the
scalar spectrum contains five physical states: two $CP$-even particles
$h,H$, the $CP$-odd scalar $A^0$ and two charged states
$H^{\pm}$.
The scalar sector can be
described in terms of six independent parameters, which can be taken to
be the four masses and two mixing angles $\alpha$ and $\beta$, with
$\tan \beta = v_2 / v_1$.

Let us now consider the signatures of the $CP$-odd scalar at the LHC. If
the $A^0$ is light, it will mostly decay into the heaviest fermion pair
available. At a hadron collider, this implies that one has to rely on
rare decay modes, such as the two-photon mode with a typical branching
ratio of $10^{-3}$. As the $A^0$ gets progressively heavier, other
channels open up, notably $t\bar{t}$ and channels that contain other
scalars: $A^{0}\rightarrow Zh,\;ZH,\;W^{\pm}H^{\mp}$.
These decays have branching ratios which are typically of the order of
10\% or larger; this means that, if the $A^0$ is heavy, these channels
are likely to represent the best way for us to search for it. We will
thus consider signatures which involve the two-photon decay of the $A^0$
or the decay into a $Z$ boson and another neutral scalar, which in turn
will be detected through its decay to two photons.

\begin{figure}
\epsfxsize 2.2in
\vspace{-0.6in}
\centerline{\epsffile{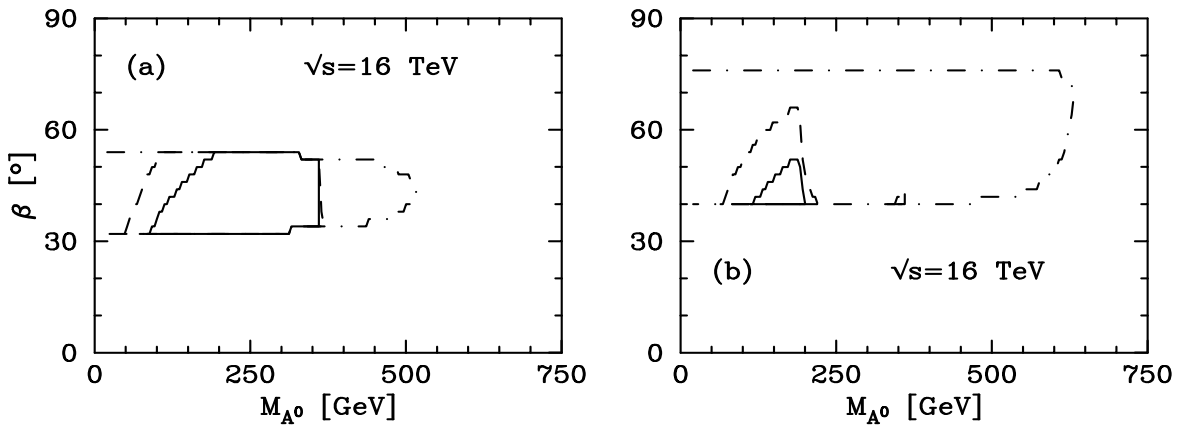}}
\caption{Significance contours for the two-photon signal, for
(a) Set~A and (b) Set~B of parameters, as explained in the text. The
outer curve represents the triviality bound. The statistical
significance is at least 5$\sigma$ in the interior of the
contours. Results are shown for integrated luminosities of
10~fb$^{-1}$ (solid) and 100~fb$^{-1}$ (dashed line).
Model~I is assumed; the contours for Model~II
are similar.}
\label{gg}
\end{figure}

We illustrate our results for the inclusive two-photon signal
in Fig.~1, where, for integrated luminosities
of 10~fb$^{-1}$ and 100~fb$^{-1}$,
we show the regions of the $(\Ma, \beta)$ plane
where the statistical significance of the signal
(defined by $S/\sqrt{B}$) is greater than 5. The sections of the
parameter space depicted in this figure
correspond to the following choice of parameters:
$M_{H}=400\; {\rm GeV},\;
M_{h}=260\; {\rm GeV},\; \Mg=350\; {\rm GeV},\;
M_{t}=180\; {\rm GeV},\; \alpha=30^{o}$ for Fig.~1a, and
$M_{H}=400\; {\rm GeV},\;
M_{h}=100\; {\rm GeV},\;  \Mg=350\; {\rm GeV},\;
M_{t}=180\; {\rm GeV},\;  \alpha=-75^{o}$ for Fig.~1b. We shall refer
to these two choices as Set~A and Set~B respectively.
The irreducible background was calculated in a mass bin $\Delta M_{\gamma
\gamma} =
3\% M_{\gamma \gamma}$,
while the reducible jet background was assumed to amount to 25\% of the
direct di-photon level.
The following rapidity and transverse momentum
cuts were employed: $|\eta^{\gamma}|<2.5$ and $p_{T}^{\gamma}>20\;
{\rm GeV}$.
Further details
of the calculation can be found
in Ref.~5. Fig.~1 also illustrates the effect of the
various parameters on the triviality bounds. Note that the region of
large $\beta$ is not efficiently explored in this channel. In general, we
find that this mode can only be helpful if $\beta \simlt 60^o$.
Because of the uncertainty over the extent to which the jet backgrounds
can be contained, we also examined the process \cite{gun-marc} where the
$CP$-odd Higgs is produced in association with a $t\bar{t}$ pair and
decays to two photons, while one of the $t/\bar{t}$ decays leptonically.
Despite the fact that the lepton tag eliminates a great deal of the jet
backgrounds, it turns out that the signal itself is weak and that this
mode cannot be very helpful.

\begin{figure}
\epsfxsize 2.2in
\vspace{-0.6in}
\centerline{\epsffile{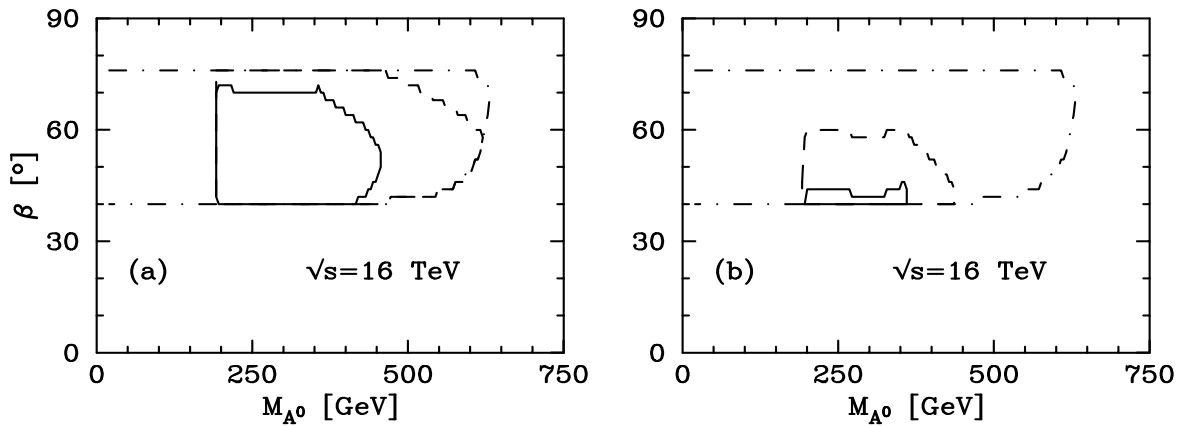}}
\caption{10-event contours for the $Z\gamma \gamma $ signal,
in the region of
the $(\Ma, \beta)$ plane allowed by triviality (outer curve)
in (a) Model~I and (b) Model~II. Set~B of parameters is used.
More than 10 events are expected in the interior of the contours.
Results are shown for integrated luminosities of
10~fb$^{-1}$ (solid) and 100~fb$^{-1}$ (dashed line).}
\label{zgg}
\end{figure}

In contrast, the process $A^0 \rightarrow Z(\rightarrow l^+l^-)\,
h(\rightarrow \gamma \gamma )$ (where $l = e,\mu $) can give
significant signals, provided 40~GeV $\simlt M_h \simlt $ 160~GeV.
Tens or even hundreds of $Z\gamma\gamma$ events can be accumulated
if $|\alpha |$ is large in Model~I or small in Model~II.
We have assumed the same rapidity and $p_T$ cuts as in the $\gamma
\gamma$ case; furthermore, we have imposed an isolation cut of
$\Delta R > 0.4$ on all pairs of final state particles.
Fig.~2 shows contours of the regions where 10 events or more are
expected over a background which turns out to be truly
negligible \cite{cpodd}. This plot corresponds to Set~B of parameters,
as defined
above, and illustrates the fact that with this mode one can explore regions of
parameter space which are inaccessible to the two-photon mode. These are
generally regions of large $M_{A^0}$, as in Fig.~2, but they can also
be regions of relatively large $\beta$.


In conclusion, we examined three different signatures of the $CP$-odd
scalar $A^0$ of the ($CP$-conserving) \THDM\ at the LHC: (i) the
two-photon decay appears to be promising if $M_{A^0} < 2\,M_t$ and
$M_{A^0} < M_h+M_Z$ and for not too large values of $\beta$; (ii) the
associated $t\bar{t}A^0$ production followed by $A^0\rightarrow \gamma
\gamma $ and tagged by the leptonic decay of the $t$ or $\bar{t}$ proved
to give a very weak signal throughout the parameter space under
investigation; and (iii) the decay $A^0 \rightarrow Zh$ followed by
$h\rightarrow \gamma \gamma,\; Z \rightarrow l^+l^-$ gives rise to
clear signals in substantial regions of parameter space, in particular
regions which are largely complementary to those covered by the other
modes examined.

This work was supported in part under NSF contract
PHY-9057173 and DOE contract DE-FG02-91ER40676.

\end{document}